\documentclass[a4paper,11pt]{article}
\pdfoutput=0

\usepackage{jcappub}
\usepackage[T1]{fontenc} 

\usepackage{subcaption}

\title{\boldmath Probing the astrophysical origin of high-energy cosmic-ray electrons with Monte Carlo simulation}

\author{R. Attallah}
\affiliation{Badji Mokhtar University, Physics Department,\\P. O. Box 12, Annaba 23000, Algeria}
\emailAdd{reda.attallah@univ-annaba.dz}

\abstract{High-energy cosmic-ray electrons reveal some remarkable spectral features, the most noteworthy of which is the rise in the positron fraction above 10~GeV. Due to strong energy loss during propagation, these particles can reach Earth only from nearby sources. Yet, the exact nature of these sources, which most likely manifest themselves in the observed anomalies, remains elusive. The many explanations put forward to resolve this case range from standard astrophysics to exotic physics. In this paper, we discuss the possible astrophysical origin of high-energy cosmic-ray electrons through a fully three-dimensional time-dependent Monte Carlo simulation. This approach, which takes advantage of the intrinsic random nature of cosmic-ray diffusive propagation, provides valuable information on the electron-by-electron fluctuations, making it particularly suitable for analyzing in depth the single-source scenario.}

\begin{document}
\maketitle
\flushbottom

\section{Introduction}
\label{sec:introduction}

Cosmic rays are high-energy charged particles striking Earth from all directions since time immemorial. They originate mainly in outer space and comprise roughly 99\% of atomic nuclei and 1\% of electrons. They are commonly divided into primary and secondary cosmic rays. Primary cosmic rays, such as electrons, protons, helium, iron and other nuclei synthesized in stars, consist of particles directly accelerated at sources. Secondary cosmic rays, like lithium, beryllium and boron, which are not created by stellar nucleosynthesis, are composed of particles produced by the primary cosmic rays during their propagation in the interstellar medium. A very small fraction of cosmic rays are stable particles of antimatter: positrons and anti-protons. Whether or not cosmic-ray positrons and anti-protons are pure secondaries is still an open question.

Even though electrons\footnote{Unless stated otherwise, we hereafter refer to both e$^-$ and e$^+$ as simply \textit{electrons}.} represent only a tiny part of cosmic rays, they are of great interest, especially at high energy. Indeed, above a few GeV these particles are subject to severe radiative energy losses, mainly by synchrotron radiation in magnetic fields and by inverse Compton scattering in radiation fields \cite{JON65, BLU70}. These processes are so drastic that high-energy cosmic-ray electrons (HECREs) cannot travel far distances from their sources. In stark contrast to hadrons, they can reach Earth only from sources in the local neighborhood \cite{COW79}. Therefore, they can be used as a powerful probe into local cosmic-ray accelerators, which is crucial to the long-standing problem of the origin of cosmic rays. HECREs are also very important to X-ray and $\gamma$-ray astronomies, to dark matter investigation and to many other issues in high-energy astrophysics. 

Researchers have taken a keen interest in cosmic-ray electrons for a very long time. The first direct observation of these particles was achieved in 1961 with balloon-borne experiments \cite{EAR61, MEY61}. During the following years many balloon flights using different detectors were carried out \cite{DAN65, SMI68, RUB68, SCH71}. These early experiments measured the flux of cosmic-ray electrons up to several hundred GeV. Experiments then explored higher energy region using more sophisticated devices with a larger geometrical factor, longer exposure and higher proton rejection power. These detectors were chiefly of two kinds: magnetic spectrometers and emulsion chambers. The first category observed negative electrons and positrons separately up to a few tens of GeV. Such was the case with the balloon-borne experiments CAPRICE \cite{BOE00, BOE01}, HEAT \cite{DUV01, BEA04} and MASS-91 \cite{GRI02}, in addition to AMS-01, which was flown on the space shuttle Discovery in 1998 \cite{AGU07}. The second category lacked the ability of discriminating between negative electrons and positrons. However, it had the merit of extending the energy spectrum measurements far beyond the range accessible to the first category. Examples include the balloon-borne experiments ECC \cite{KOB12}, BETS \cite{TOR01}, ATIC-2 \cite{CHA05} and PPB-BETS \cite{YOS08}. Moreover, the ground-based Cerenkov telescopes H.E.S.S. \cite{AHA08} and MAGIC \cite{BOR11} also observed cosmic-ray electrons at very high energy. The launch over the last decade of a new generation of high-precision instruments on-board satellites (PAMELA, the Fermi Gamma-ray Space Telescope, AMS-02 and CALET) opened a new era in the study of HECREs. 

The detection of HECREs implies the existence of local sources. These sources are expected to manifest themselves in the energy spectrum, which should display special features. Recent experimental results show that the energy spectrum of cosmic-ray electrons extends well beyond 1~TeV and does have features. Most notably, PAMELA uncovered in the energy range 10-300~GeV a significant increase in the positron fraction (ratio of the positron flux to the combined flux of positrons and negative electrons) \cite{ADR09, ADR10, ADR13}. This remarkable result, confirmed by Fermi-LAT \cite{ACK12} and by AMS-02 \cite{AGU13}, is not consistent with conventional models based on the assumption that positrons arise only from the secondary production of cosmic rays by collision with the interstellar medium. These models rather predict a positron fraction falling smoothly with energy \cite{PRO82, MOS98}. A new precision measurement by AMS-02 extending up to 500~GeV indicates that the positron fraction, on the one hand, levels off at about 200~GeV and, on the other hand, does not show any fine structure or sharp cutoff \cite{ACC14}.

In parallel, ATIC reported an excess of cosmic-ray electrons over conventional model expectations between 300 and 800~GeV followed by a steepening above 1~TeV \cite{CHA08, PAN11}. Although PPB-BETS also showed a bump-like structure between 100 and 700~GeV \cite{YOS08}, Fermi-LAT and H.E.S.S. experiments found no evidence for a prominent peak \cite{ABD09, ACK10b, AHA09}. 

There are several possible scenarios, from standard astrophysics to exotic physics, to account for the special features of the electron energy spectrum. They may be ascribed to the contribution from some individual local sources such as supernova remnants (SNRs) (see, e.g., \cite{BLA09, AHL09, STA10}) or pulsars (see, e.g., \cite{HOO09, MAL09, YUK09, DEL10, BLA11, PRO11, DIM14}). However, additional sources such as the annihilation or decay of dark-matter particles cannot be excluded (see, e.g., \cite{KAN09, ARK09, CHO09, POR11, KOP13, BER13, IBA14, FEN14}). The main idea of these models is that equal amounts of negative electrons and positrons are produced by the source, be it an astrophysical object or dark-matter particles. This contribution emerges at high energy from a background formed by electrons coming from distant sources. The observed anomalies may also be due to just propagation effects \cite{SHA09,GAG13,BLU13,COW14}. It must be emphasized here that there are arguments for and against each scenario and none of these approaches is yet conclusive \cite{MOS13}. Recent reviews can be found in \cite{FAN10, SER12, CHO13,PAN13,PIC14,ISR14}.

The energy spectrum of cosmic-ray electrons is typically interpreted within the context of propagation models and the traditional method consists in solving appropriate transport equations \cite{GIN76b, STR07}. Due to its inherent stochastic nature, the diffusive propagation of cosmic rays can also be treated using Monte Carlo simulation \cite{HUA07}. While it is widely believed that this technique is inefficient in this kind of application, the proximity of sources in the case of HECREs, coupled with the absence of hadronic interactions and the simplicity in energy loss processes, simplifies the modeling significantly. Monte Carlo approach comes in handy particularly at very high energy where only a few sources are expected to dominate the cosmic-ray electron flux. It provides very useful information about the electron-by-electron fluctuations, thus complementing the traditional method.

In this respect, we implemented a fully 3-dimensional time-dependent Monte Carlo simulation of the propagation of HECREs in our galaxy. To speed up calculations we employed MPI parallel programming on an HPC cluster system. We restricted ourselves to the energy region above 10~GeV where cosmic rays are not affected by the solar wind modulation. We considered only pure diffusion since convection and reacceleration are negligible above a few GeV \cite{DEL09}. We focused on the most natural way to explain the spectral features of HECREs, assuming that some nearby pulsars and/or SNRs are the sources of such particles. The other possible scenarios were not examined. We used a two-component model, separating the local source contribution from the distant source contribution \cite{ATO95}. The latter, which constitutes the background, was estimated with the public code GALPROP\footnote{http://galprop.stanford.edu/}  \cite{VLA11}.

There are two different types of key quantities when interpreting the observed energy spectrum of cosmic-ray electrons. The first ones are the parameters associated with propagation; they are behind the distortion of the injection spectrum. The second ones are the parameters related to the spectral profile at injection; they control the overall form of the measured spectrum. In our study we first examined the propagation effects. Next, we inferred from these initial calculations the most likely astrophysical sources of HECREs. Then, we derived the expected flux from some typical sources. And finally, we addressed the anisotropy issue.

The outline of this paper is as follows. After this introduction (\S\ref{sec:introduction}) we review in \S\ref{sec:propagation} the necessary mathematical background for describing cosmic-ray electron propagation and we detail our Monte Carlo procedure. In the following section (\S\ref{sec:results}) we report the main results achieved: electron energy and lifetime distributions in \S\ref{sec:prop_effects}, potential astrophysical sources in \S\ref{sec:sources},  cosmic-ray electron flux, positron fraction and anisotropy for some typical sources in \S\ref{sec:flux} and \S\ref{sec:anisotropy}. We end our study with a brief conclusion (\S\ref{sec:conclusion}).

\section{Propagation of cosmic-ray electrons}
\label{sec:propagation}

Many aspects of the origin of most cosmic rays can be understood in terms of acceleration at astrophysical sources and propagation in turbulent magnetic fields in the Galaxy (see, e.g., \cite{BER90,GAI90,LON11}). Because of random scattering by the magnetic field irregularities, charged cosmic-ray particles can be considered to diffuse from their sources through the interstellar medium. The principal argument in favor of diffusion comes from the presence in the cosmic radiation of a much greater proportion of secondary nuclei when compared to the elemental abundances of the solar system. Measurements of the ratios of secondary to primary cosmic-ray nuclei, such as B/C, lead to the following conclusions:
\begin{enumerate}
 \item Cosmic rays travel distances thousands of times greater than the thickness of the galactic disk between injection and observation. This result suggests diffusion in a confinement region including the galactic disk.
 \item Most of propagation occurs after acceleration since the amount of matter traversed by cosmic rays decreases as energy increases.
\end{enumerate}
\noindent Cosmic-ray electrons should not depart from the general rule and should also undergo diffusion between injection and observation.

\subsection{Energy loss processes}

As cosmic-ray electrons propagate through the interstellar medium, they are subject to a number of energy loss processes that cause distortion of their injection spectra. These processes, which involve interaction with matter, with magnetic fields and with radiation, comprise \cite{LON11}:
\begin{itemize}
 \item[--] ionization loss, which depends upon energy $E$ logarithmically ($\propto \ln E$);
 \item[--] bremsstrahlung (and adiabatic loss when relevant), which varies linearly with energy ($\propto E$);
 \item[--] synchrotron radiation in magnetic fields and inverse Compton scattering in radiation fields, which are proportional to the square of energy ($\propto E^2$).
\end{itemize}

\begin{figure}[tbp] 
  \centering
  \includegraphics[width=0.58\textwidth]{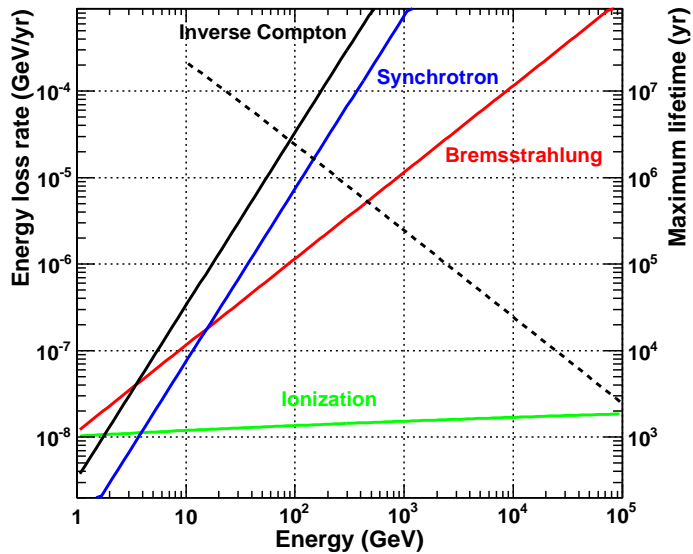}
  \caption{Energy loss rates as a function of energy of cosmic-ray electrons during their propagation in the Galaxy. The dashed line represents the evolution of the electron maximum lifetime with energy.}
  \label{fig:energy_loss}
\end{figure}

The different energy loss processes of cosmic-ray electrons are shown in figure~\ref{fig:energy_loss} for typical values of the interstellar gas density $n$ ($\simeq 1$~cm$^{-3}$), the galactic magnetic field $B$ ($\simeq 3 \, \mu$G) and the energy density of the galactic radiation field $U_{\rm rad}$ ($\simeq 1$~eV~cm$^{-3}$) \cite{LON11}. The galactic radiation is composed of the stellar radiation, the re-emitted radiation from dust grains and the cosmic microwave background (CMB). Except for very low energy, the ionization loss can be ignored. Bremsstrahlung is important in the GeV energy range. Synchrotron radiation and inverse Compton scattering predominate above a few tens of GeV, the latter being about three times stronger than the former. 

For HECREs ($\gtrsim 10$~GeV) it is quite reasonable to consider only bremsstrahlung, synchrotron radiation and inverse Compton scattering. The total energy loss rate by these processes for ultra-relativistic electrons  within Thomson approximation is given by \cite{LON11}
\begin{equation}
    -\frac{{\rm d}E}{{\rm d}t} = a_0 E + b_0 E^{2} ,  
    \label{eq:loss_rate}
\end{equation}
\noindent with
\begin{equation}
    a_0 = 3.66 \times 10^{-22} \; n \simeq 3.7 \times 10^{-16} \; \; \rm{s}^{-1} ,
    \label{eq:a0}
\end{equation}
\noindent and
\begin{equation}
    b_0 = \frac{4 \sigma_{\rm T} c}{3 (m_{\rm e}c^{2})^{2}} \left( U_\mathrm{mag} + U_\mathrm{rad} \right) 
    \simeq 1.3 \times 10^{-16} \; \; \rm{ GeV}^{-1} \; \rm{s}^{-1} .
    \label{eq:b01}
\end{equation}
\noindent $\sigma_{\rm T}$ is Thomson scattering cross-section, $c$ the speed of light, $m_{\rm e}$ the rest mass of the electron and $U_\mathrm{mag} (\equiv B^2 / 2 \mu_0)$ the energy density of the galactic magnetic field ($\mu_{0}$ is the permeability of free space). Because of these energy losses, the maximum lifetime of a cosmic-ray electron with an energy $E$ anywhere in the Galaxy is therefore
\begin{equation}
    \tau_{\rm max} = \int_E^\infty \frac{{\rm d}E}{(-{\rm d}E/{\rm d}t)} = \frac{1}{a_0} \ln \left( \frac{a_0+b_0 E}{b_0 E} \right) .
    \label{eq:tau}
\end{equation}
\noindent Above a few tens of GeV bremsstrahlung becomes negligible and then
\begin{equation}
    \tau_{\rm max} \simeq \frac{1}{b_{0}E} \qquad \left( E \gg \frac{a_0}{b_0} \simeq 3 \, \, \, \mathrm{GeV} \right) .
    \label{eq:tau2}
\end{equation}
\noindent $\tau_{\rm max}$ decreases very rapidly with energy (figure~\ref{fig:energy_loss}). For 10, $10^2$ and $10^3$~GeV, $\tau_{\rm max}$ is roughly equal to $10^7$, $10^6$ and $10^5$~yr, respectively.

Actually, at very high energy the constant $b_0$ should be corrected according to the full Klein-Nishina cross-section for the inverse Compton scattering interactions of electrons. This correction results in a significant reduction of the electron energy loss rate compared with Thomson approximation. It can be formulated as \cite{SCH10}:
\begin{equation}
  b_0 = \frac{4 \sigma_{\rm T} c U_\mathrm{mag}}{3 (m_{\rm e}c^{2})^{2}} \left( 1 + \sum_{i=1}^4 \frac{U_\mathrm{rad}^i}{U_\mathrm{mag}} \frac{E_{K,i}^2}{E^2+E_{K,i}^2} \right) .
  \label{eq:b02}
\end{equation}
\noindent The sum relates to the four types of radiation existing in the Galaxy: optical photons from stars of spectral type B, optical photons from stars of spectral type G-K, infrared photons and CMB photons. $U_\mathrm{rad}^i$ and $E_{K,i}$ ($i = 1$-4) are the corresponding energy densities and critical Klein-Nishina energies, respectively. Their values can be found in ref. \cite{SCH10} (Table~1).

\subsection{Description of diffusion}

Diffusion is experienced as a macroscopic phenomenon although the mechanisms of diffusion are founded in microscopic collision processes. The phenomenological theory of cosmic-ray diffusion leads to a well-known transport equation, called the diffusion-loss equation \cite{GIN64}. In the case of homogeneous and isotropic diffusion, it is given for electrons by
\begin{equation}
  \frac{\partial \phi(E, \, \mathbf{r}, \, t)}{\partial t} = D(E) \nabla^2 \phi (E, \, \mathbf{r}, \, t)
  + \frac{\partial }{\partial E} \left[ b(E) \phi (E, \, \mathbf{r}, \, t) \right] + Q(E, \, t) ,
  \label{eq:transport}
\end{equation}
\noindent where $\phi (E, \, \mathbf{r}, \, t)$ is the density of cosmic-ray electrons with energy between $E$ and $E + \mathrm{d} E$ at position $\mathbf{r}$ and time $t$, $D(E)$ the energy-dependent diffusion coefficient, $Q(E, t)$ the source term, i.e. the rate of injection of cosmic-ray electrons, and $b(E)$ the mean rate at which electrons loose energy:
\begin{equation}
  b(E) = -\frac{\mathrm{d} E}{\mathrm{d} \, t} .
  \label{eq:be}
\end{equation} 

The diffusion coefficient $D$ sets the level of turbulence that scatters cosmic-ray charged particles. It is linked to the diffusion mean free path $\lambda$ \cite{GAI90}:
\begin{equation}
  D = \frac{1}{3} \lambda v_{\parallel} ,
  \label{eq:d}
\end{equation}
\noindent where $v_{\parallel}$ is the projection of the electron velocity onto the magnetic line direction along which the electron moves. As long as the magnetic field is slowly evolving, cosmic-ray electrons move along the same field lines. However, the random scattering by the magnetic field inhomogeneities results in a uniform distribution of pitch angles over the range [0, $\pi/2$] \cite{LON11}. For relativistic electrons, the quantity $v_\parallel$ is then a random number between 0 and $c$, depending on the value of the pitch angle.

One of the main sources of uncertainty in diffusive propagation models comes from the diffusion coefficient $D$, which is calculated by using the observed ratios of secondary to primary cosmic-ray nuclei (e.g. B/C) in the GeV energy region. But the uncertainties are still large and, moreover, there is some lack of experimental data in the TeV energy region, which makes it difficult to discriminate among different diffusion coefficients. A widely adopted expression for relativistic particles is given by
\begin{equation}
  D(E) = D_0 (E/\mathrm{GeV})^\delta ,
  \label{eq:de}
\end{equation}
\noindent where $\delta = 0.3$-0.6 and $D_0 = (1$-5)$\times$10$^{28}$ cm$^2$ s$^{-1}$ \cite{STR07}. Accordingly, the diffusion mean free path of cosmic-ray electrons also depends on energy:
\begin{equation}
  \lambda(E) = \frac{\lambda_0 (E/\mathrm{GeV})^\delta}{\beta_{\parallel}},
  \label{eq:lambda}
\end{equation}
\noindent where $\beta_{\parallel} = v_{\parallel}/c$ and $\lambda_0 = 3 D_0 / c \simeq 1$~pc.

A simple treatment of diffusion (random walk) shows that the root mean square (RMS) of individual displacements, i.e. the average straight-line distance between the origin and the position of particles after time $t$, is given by
\begin{equation}
\sqrt{\langle r^2 \rangle} = \sqrt{6 D t} .
  \label{eq:distance}
\end{equation}
\noindent Replacing $t$ with the maximum lifetime ($\tau_\mathrm{max}$) in this purely phenomenological relation, one can readily estimate the maximum distance of cosmic-ray electron sources. For $10$, $10^2$ and $10^3$~GeV, these distances are roughly equal to 7, 3 and 2~kpc, respectively. Thus, only very local sources within a few kpc from the solar system should contribute effectively to the electron spectrum at high energy.

\subsection{Monte Carlo approach}

Astrophysical sources are discrete and time-dependent, unlike dark-matter sources, which are supposed to be continuously distributed and independent of time. The electron acceleration at astrophysical sources, such as pulsars, lasts only a few tens of kyr, while the time scale (typical propagation time) is of the order of several hundred of kyr. Moreover, the sources are typically a few pc in size, while the space scale (typical source distance) is about several hundred pc. As a result, one can use the so-called burst-like approximation \cite{SHE70}, which regards astrophysical sources as point-like and instantaneous. Nonetheless, this approximation is not a crucial element of our model and a delay in electron emission with respect to the source birth can be handled quite easily. We just have to deduct this delay from the arrival time of each observed electron without the need to redo the calculations. 

The calculation starts from the selected astrophysical source, where one electron is picked with an energy $E$. Unless $E$ is fixed (monochromatic source), the energy injection spectrum is sampled for more efficiency with the ziggurat algorithm \cite{MAR00}. Assuming homogeneous and isotropic diffusion, the electron is injected into space with a random three-dimensional direction. The path length $l$ is generated according to the probability distribution
\begin{equation}
    P(l) \propto \exp \left( -l / \lambda \right) ,
    \label{eq:P}
\end{equation}
\noindent where $\lambda$ is calculated with equation (\ref{eq:lambda}).  Between two scatterings, the electron is supposed to move along a straight path because the Larmor radius ($\sim 10^{-5}$~pc) is considerably smaller than the mean free path ($\sim 1$~pc). After traveling the distance $l$, the electron energy is adjusted taking into account the energy loss rate given by relation (\ref{eq:loss_rate}). Using the approximation of a continuous energy loss, the new energy $E'$ is given by
\begin{equation}
    E' = \frac{a_0 E}{\exp (a_0l/v_{\parallel}) (a_0 + b_0 E) - b_0 E} .  
    \label{eq:new_energy}
\end{equation}
\noindent The elapsed time is calculated by taking the instant of injection as the time origin, that is an instantaneous release of electrons is supposed. The electron continues to propagate and the whole process is iterated until one of these three events occurs:
\begin{enumerate}
  \item The electron reaches the boundaries of the confinement region, in which case it is discarded (free escape) and a new electron is simulated starting from the source;
  \item The electron does not reach the solar system within a maximum time (=~2$\times$10$^7$~yr) corresponding to a minimum simulation energy of 10~GeV, in which case it is also discarded and the next electron is considered;
  \item The electron crosses a shell of 1~pc radius centered on the Sun, in which case the electron is considered to be observed (or detected) and all its characteristics (energy, lifetime and direction) are recorded.
\end{enumerate}

The geometry of the confinement region is fixed according to the flat halo diffusion model \cite{GIN76a}, which presumes that the Galaxy has the shape of a cylindrical slab with a radius $R$ and total height $2h$. In our case $R = 20$~kpc and $h = 4$~kpc. The distance of the Sun to the galactic center is fixed at 8.5~kpc as usual.

It is obvious that the detection shell should be as small as possible to be considered as point-like (the size of the solar system is less than one thousandth of a pc). However, the smaller the detection shell the smaller the detection efficiency and the larger the computational time. Since the mean free path of cosmic-ray electrons above 10~GeV is greater than 2~pc (\ref{eq:lambda}), the value of 1~pc used here for the radius of the detection shell is a good trade-off. This requirement ensures that no significant diffusion will occur within the detection shell.

We adopted throughout our analysis intermediate values for the diffusion coefficient ($\delta = 0.45$, $D_0 = 3 \! \! \times \! \! 10^{28}$ cm$^2$ s$^{-1}$), and mean values of the interstellar gas density (=~1~cm$^{-3}$) and the galactic radiation field density (=~1~eV~cm$^{-3}$). In addition, we only considered the average value of the galactic magnetic field (=~3~$\mu$G) to keep the model as simple as possible in its first version and save computing time. The fluctuations of the galactic magnetic field undoubtedly play a key role (see, e.g., \cite{HAR16, KIS12}); they will be the subject of a follow-up dedicated study.

For this large scale simulation program developed to run on parallel systems, the quality of the pseudo-random number generator is a crucial factor. We used Tina's Random Number Generator Library \cite{BAU14}, which is especially suited for parallel programming environment. To generate independent  streams of random numbers on each processing node, we adopted the leapfrog parallelization technique. Typical calculations take 30-50~hours with 128 cores running at full speed.

\section{Results and discussion}
\label{sec:results}

\subsection{Propagation effects}
\label{sec:prop_effects}
\begin{figure*}[t!]
  \centering
  \begin{subfigure}[t]{0.5\textwidth}
    \centering
    \includegraphics[width=0.9\textwidth]{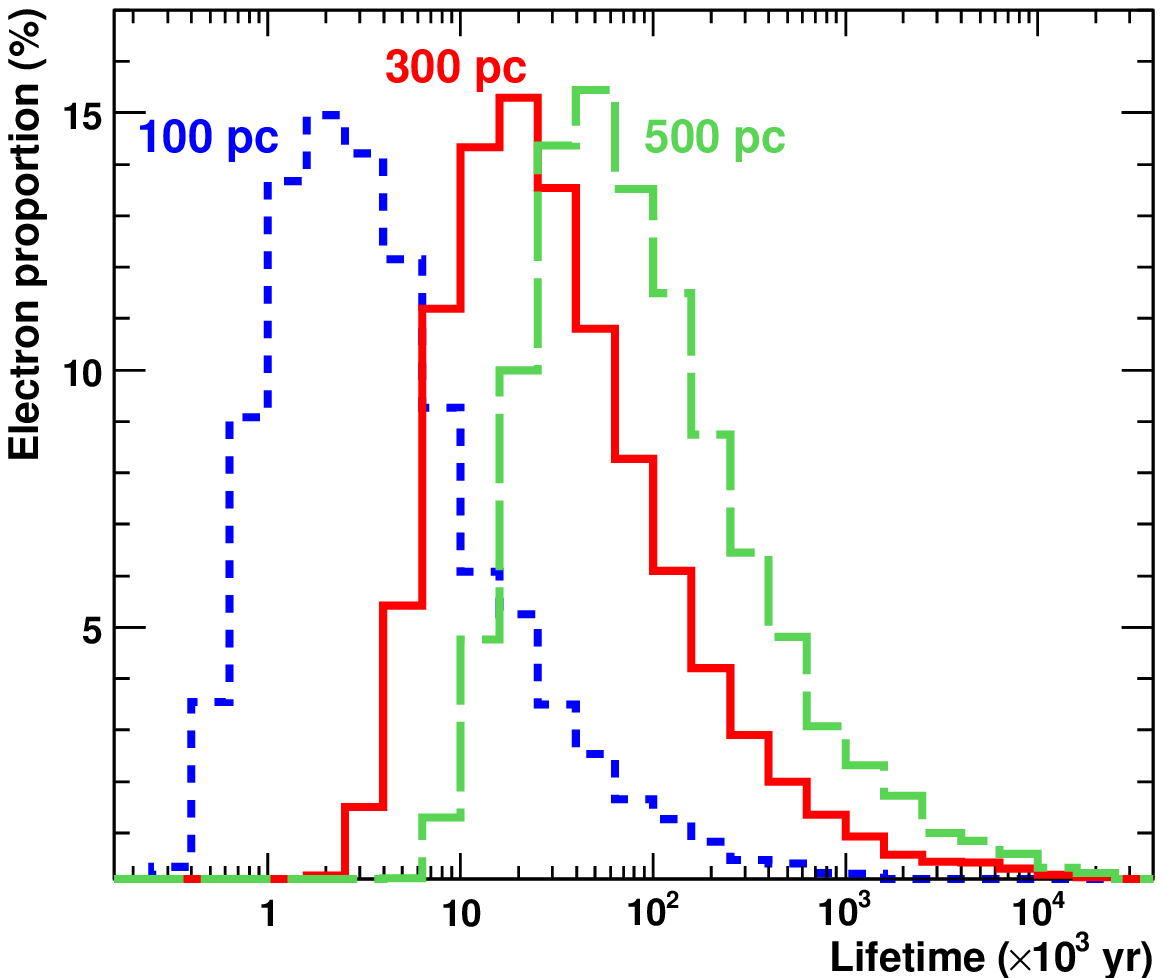}
    \caption{}
    \label{fig:time1}
  \end{subfigure}%
  ~
  \begin{subfigure}[t]{0.5\textwidth}
    \centering
    \includegraphics[width=0.9\textwidth]{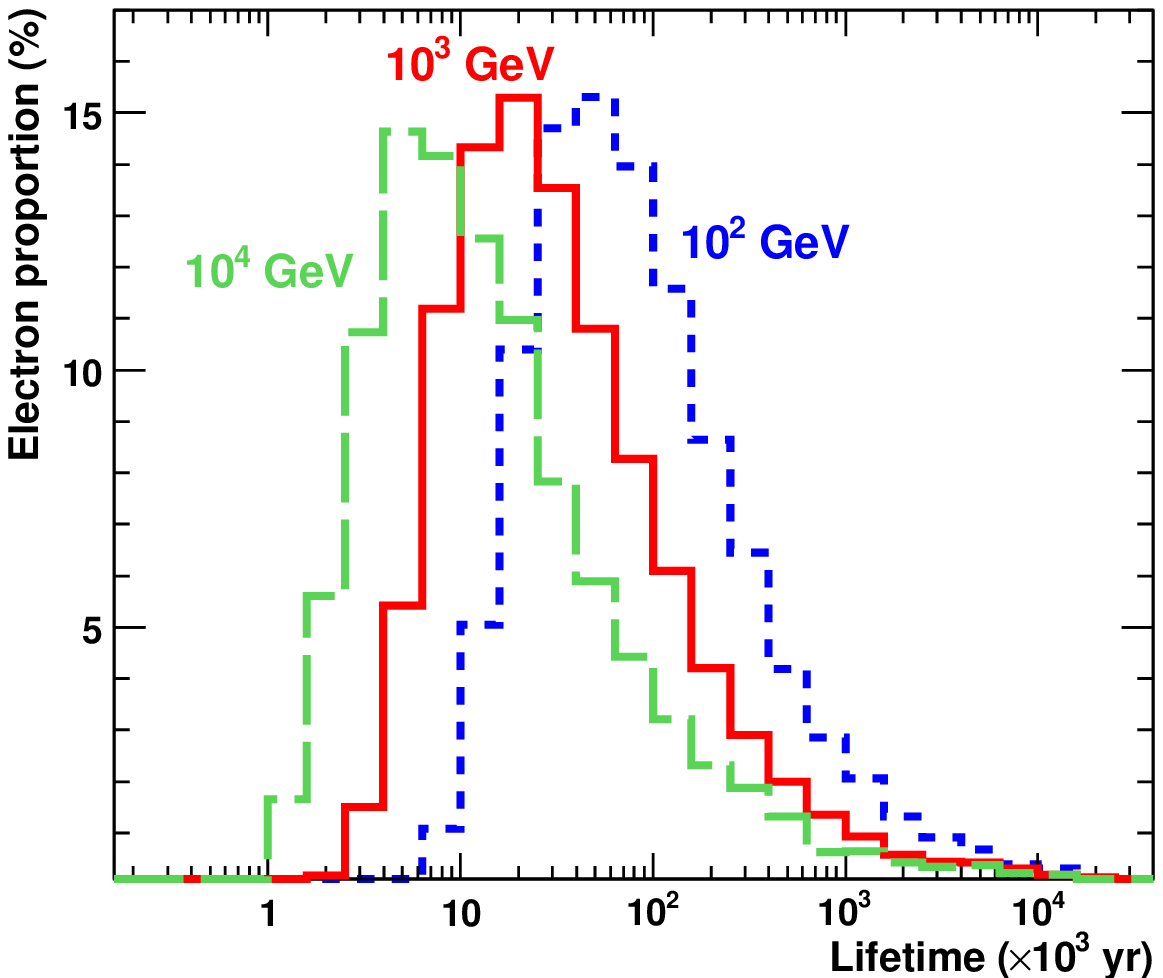}
    \caption{}
    \label{fig:time2}
  \end{subfigure}
  \caption{Lifetime distributions of observed cosmic-ray electrons for: (a)~a monochromatic source ($10^3$~GeV) located at different distances from the solar system; (b)~different monochromatic sources located at the same distance (300~pc) from the solar system.}
\label{fig:time}
\end{figure*}

\begin{figure*}[t!]
  \centering
  \begin{subfigure}[t]{0.5\textwidth}
    \centering
    \includegraphics[width=0.9\textwidth]{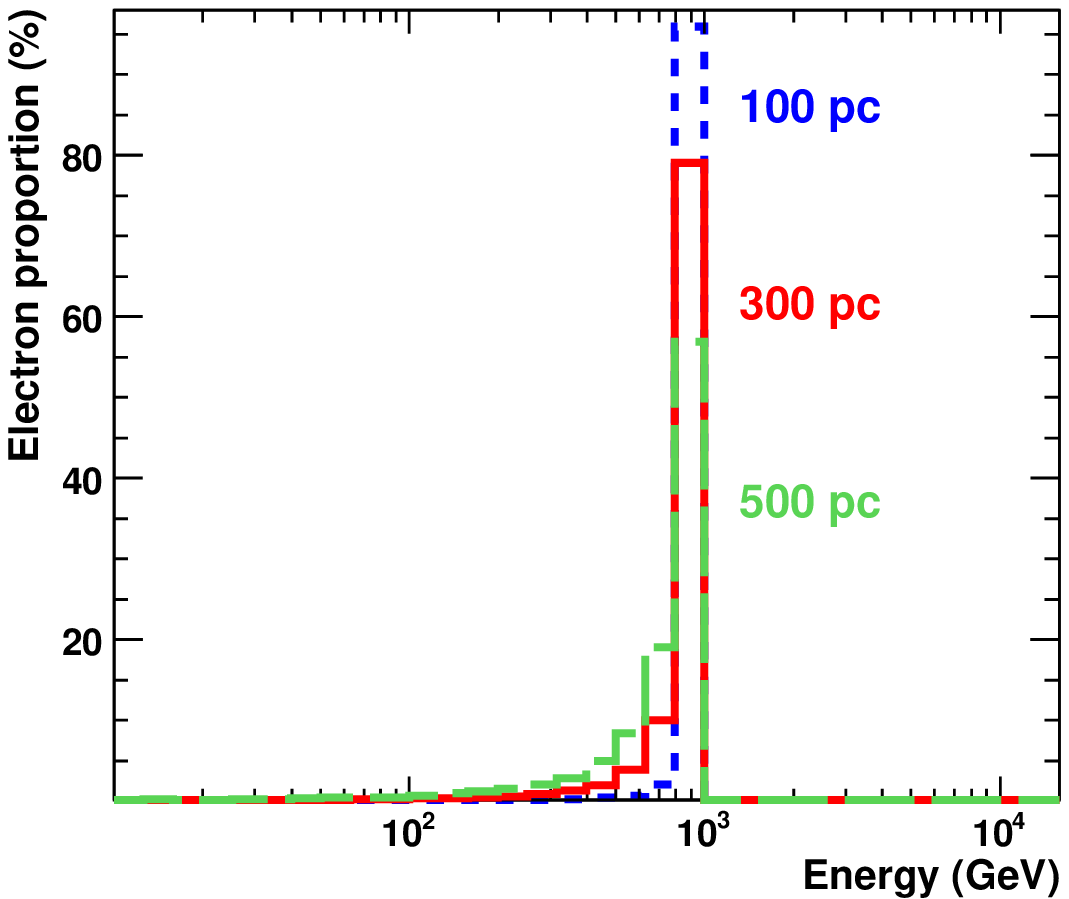}
    \caption{}
    \label{fig:energy1}
  \end{subfigure}%
  ~
  \begin{subfigure}[t]{0.5\textwidth}
    \centering
    \includegraphics[width=0.9\textwidth]{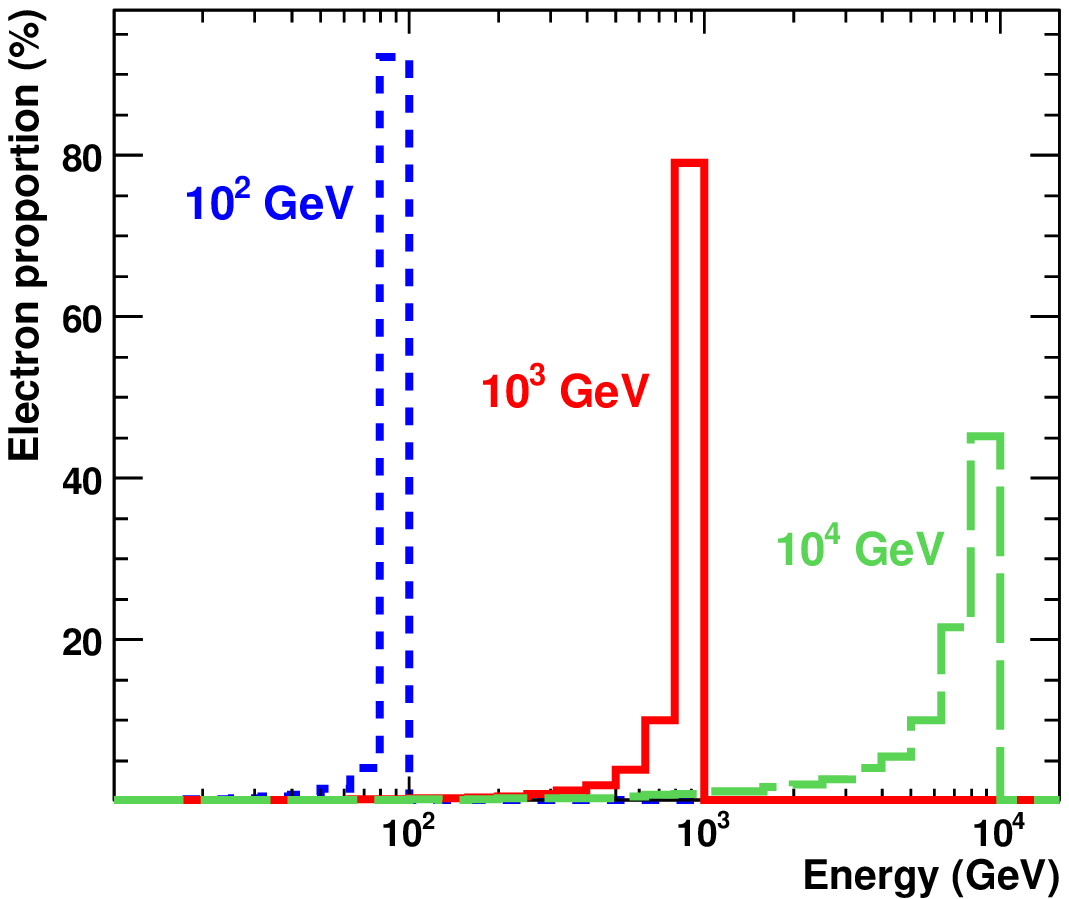}
    \caption{}
    \label{fig:energy2}
  \end{subfigure}
\caption{Energy distributions of observed cosmic-ray electrons for: (a) a monochromatic source ($10^3$~GeV) located at different distances from the solar system; (b)~different monochromatic sources located at the same distance (300~pc) from the solar system.}
\label{fig:energy}
\end{figure*}

To assess the impact of propagation we first considered monochromatic electron sources located at different distances from the solar system (with null values for the galactic longitude and latitude). We calculated the lifetime and energy distributions of the observed electrons for different values of the injected energy. Examples of such distributions are shown in figure~\ref{fig:time} and \ref{fig:energy}. The calculations were carried out so as to have $10^4$ electrons crossing the detection shell in each run, which corresponds to a total number of simulated events (electrons) in the range $10^8$-$10^9$ per run. It is clear that the farther the source the greater the number of simulated events needed to meet this requirement.

As indicated in figure~\ref{fig:time}, the lifetime distributions are characterized by a dramatic rise followed by a steady decline with a very long tail. Please note that the time scale is logarithmic. The distributions are severely right-skewed (skewness $\sim +12$ for the 300~pc source). Around 50\% of the electrons reach the solar system quite quickly, within a few tens of kyr. The rest arrive later over a much longer period ($\sim 10^4$~kyr). When the distance of the source increases, the lifetime distribution as a whole moves to the right because electrons spend more time when traveling from farther sources (figure~\ref{fig:time1}). On the contrary, when the initial energy increases, the lifetime distribution moves to the left (figure~\ref{fig:time2}). Since the mean free path increases with energy ($\sim E^\delta$), higher-energy electrons spend less time when diffusing from the same source to Earth. However, this effect is partly counterbalanced by the energy loss mechanisms, which are stronger for higher-energy electrons.

We have quite the opposite with the energy distributions shown in figure~\ref{fig:energy}. These distributions grow very slowly up to the maximum available energy then fall off very steeply. They are also highly left-skewed (skewness $\sim -3$ for the 300~pc source). The higher-energy part of the spectrum involves the electrons reaching the solar system early, whereas the lower-energy part involves the electrons arriving late. The latter have diffused in the confinement region a much longer time and thus have lost more energy.  When the distance of the source increases, the energy distribution flattens because electrons undergo more energy loss when they originate from farther sources (figure~\ref{fig:energy1}). On the other hand, when the initial energy increases, the energy distribution naturally moves to the right (figure~\ref{fig:energy2}). But it also flattens because of the energy loss processes, which are harder at higher energy.

We can deduce from these initial calculations the following:
\begin{itemize}
 \item[--] The electron lifetime distribution depends strongly on the source distance and the initial energy; 
 \item[--] Assuming the burst-like approximation, there exists a ``right'' timing of electron emission for each position of the source, leading to the most effective contribution to the electron spectrum. Indeed, the contribution is more significant if the source age is around the time corresponding to the peak of the lifetime distribution;
 \item[--]  Sources with ages beyond the time at the peak of the lifetime distribution also contribute owing to the strong right-skewness of the distributions, but much less significantly; Otherwise, if the age is far below the distribution peak time, the solar system will receive at best a very small fraction of electrons (most electrons haven't arrived yet).
\end{itemize}

\subsection{Possible astrophysical sources}
\label{sec:sources}
It is commonly believed that cosmic rays up to $\sim 10^{15}$~eV, including electrons, are accelerated through diffusive shock acceleration in SNRs \cite{ELL07}. The observation of X-ray synchrotron emission in SNRs, such as SN~1006 \cite{KOY95} and Cassiopeia A \cite{ALL97}, provides compelling evidence for the presence of high-energy electrons. The detection of TeV $\gamma$-ray emission from SN~1006 \cite{TAN98} and G347.3-0.5 \cite{MUR00}, for instance, corroborates the acceleration of electrons by SNRs to about 100~TeV. So it is quite natural to invoke these objects to interpret the cosmic-ray electron spectrum at high energy \cite{KOB01, ERL02, KOB04}. However, if SNRs can accelerate not only negative electrons but also positrons, the ratios B/C and $\bar{\mathrm p}$/p should increase with energy \cite{CHO14}. Such a rise has simply not yet been observed experimentally, with the exception of a very recent observation of an anti-proton excess by AMS-02 \cite{KOU15}.

Furthermore, the idea that the spectrum of cosmic-ray electrons is dominated at very high energy by the contribution from a few local pulsars or pulsar wind nebulae (PWN) has long been known \cite{SHE70, BOU89, AHA95}. These rapidly spinning magnetized neutron stars are believed to be excellent factories of electron-positron pairs. The underlying mechanism is that electrons are first extracted from the surface by the intense electric field induced by the rotation of the star. Then they are transformed into electron-positron pairs through electromagnetic cascades in the pulsar magnetosphere \cite{REE74, ZHA01, AMA14}. Pulsars provide a natural explanation to the peculiar features observed in the cosmic-ray electron spectrum at high energy, especially the rise in the positron fraction. However, the way electrons escape the accelerator environment has yet to be understood \cite{BLA11}.

\begin{figure}[tbp] 
  \centering
  \includegraphics[width=0.60\textwidth]{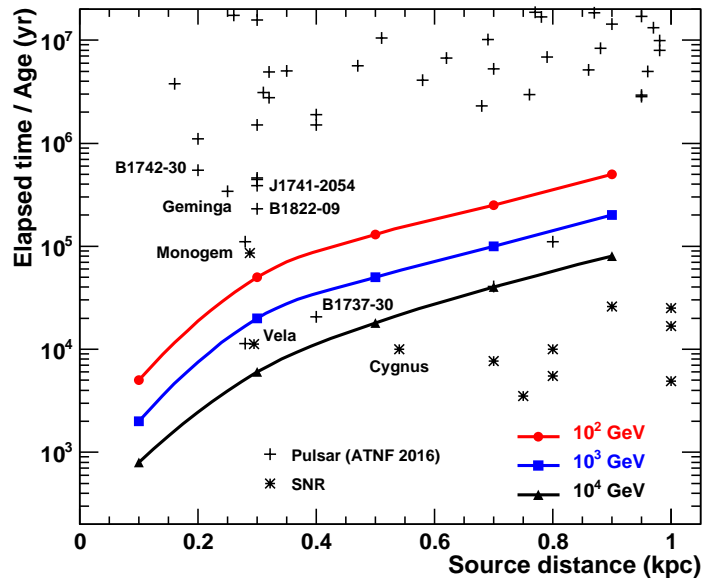}
  \caption{Relationship between the source distance and the electron elapsed time at the peak of the lifetime distribution for different values of the initial energy. Nearby pulsars and SNRs with distance and age as coordinates are superimposed on the same graphic.}
  \label{fig:sources}
\end{figure}

The curves of figure~\ref{fig:sources} graphically depict the relationship between the elapsed time at the peak of the lifetime distribution and the distance of the source for different values of the initial energy. If we make no presumption in advance about the source nature and suppose a prompt release of electrons, the most likely sources of HECREs should lie very close to these curves. To determine potential astrophysical sources, all we have to do is superimpose on the same graphics pulsars and SNRs by taking their distances and ages as coordinates. As explained previously, only nearby sources within about 1~pc and aged at most $10^7$~yr have relevance. Pulsar data are taken from the catalog of the Australia Telescope National Facility (ATNF)\footnote{http://www.atnf.csiro.au/research/pulsar/psrcat/} \cite{MAN05} and SNR data from \cite{DEL10} (Table C.1) and references therein.

As can be seen in figure~\ref{fig:sources}, Monogem and Vela are the closest ($\sim 300$~pc) among the potential sources of HECREs and are then the leading candidates. Owing their respective ages, Monogem might well be the main contributor to the electron spectrum below 100~GeV and Vela in the TeV energy range. Interestingly, these objects are SNRs with known pulsar counterparts (PSR B0656+14 and B0833-45). However, it is generally held that the produced e$^+$e$^-$ pairs are first trapped in the PWN or the SNR surrounding the source and released into the interstellar medium a few tens of kyr after the birth of the source \cite{BUS08, HOO09, PRO11}. In this sense, Vela (and PSR B1737-30) appears too young to play any role. Geminga (PSR J0633+1746) and PSR B1822-09 are the ``next-in-line`` candidates and might also contribute below 100~GeV. These results corroborate the thesis of mature SNRs/pulsars already pointed to by many earlier studies \cite{BUS08, HOO09, GRA09, LIN13, ERL13, YIN13}. Other interesting objects include PSR B1742-30 and PSR J1741-2054. The other possible sources located farther away are naturally disadvantaged since the flux obeys the inverse-square law.

A ``right'' position coupled with a ``right'' age is a necessary but certainty not sufficient condition for the sources of HECREs. The spectral profile at injection has also a major impact on the observed flux. According to the theory of cosmic ray acceleration mechanism by non-relativistic expanding shock waves in star explosions, the energy spectrum of electrons injected by the source into the interstellar medium can be expressed as a power law  with a characteristic exponential energy cutoff ($E_\mathrm{cut}$):
\begin{equation}
  Q(E) = Q_0 E^{-\gamma} \exp \left( - E / E_\mathrm{cut} \right) ,
\end{equation}
\noindent where $\gamma$ is the spectral index and $Q_0$ a normalization factor. This form can be used for SNRs as well as for pulsars. The parameters $\gamma$, $E_\mathrm{cut}$ and $Q_0$ are estimated on the basis of either experimental studies or theoretical considerations. The spectral index $\gamma$ is found to be $\gtrsim 2$ for SNRs and $\lesssim 2$ for PWN \cite{ELL07, ABD13}. The energy cutoff is not well known but there are indications that it lies in the TeV energy range \cite{AHA08, AHA06}. $Q_0$ can be determined from the relation
\begin{equation}
  \int_0^\infty Q(E) \ E \ \mathrm{d} E = \eta W,
\end{equation}
\noindent where $W$ is the total spin-down energy of the source and $\eta$ the fraction of energy transmitted into e$^-$e$^+$ pairs. There are arguments that the maximum value of $\eta W$ is on the order of $10^{48}$-$10^{50}$~erg \cite{MAL09, KOB04}.

\begin{figure*}[t!]
  \centering
  \begin{subfigure}[t]{0.5\textwidth}
    \centering
    \includegraphics[width=0.80\textwidth]{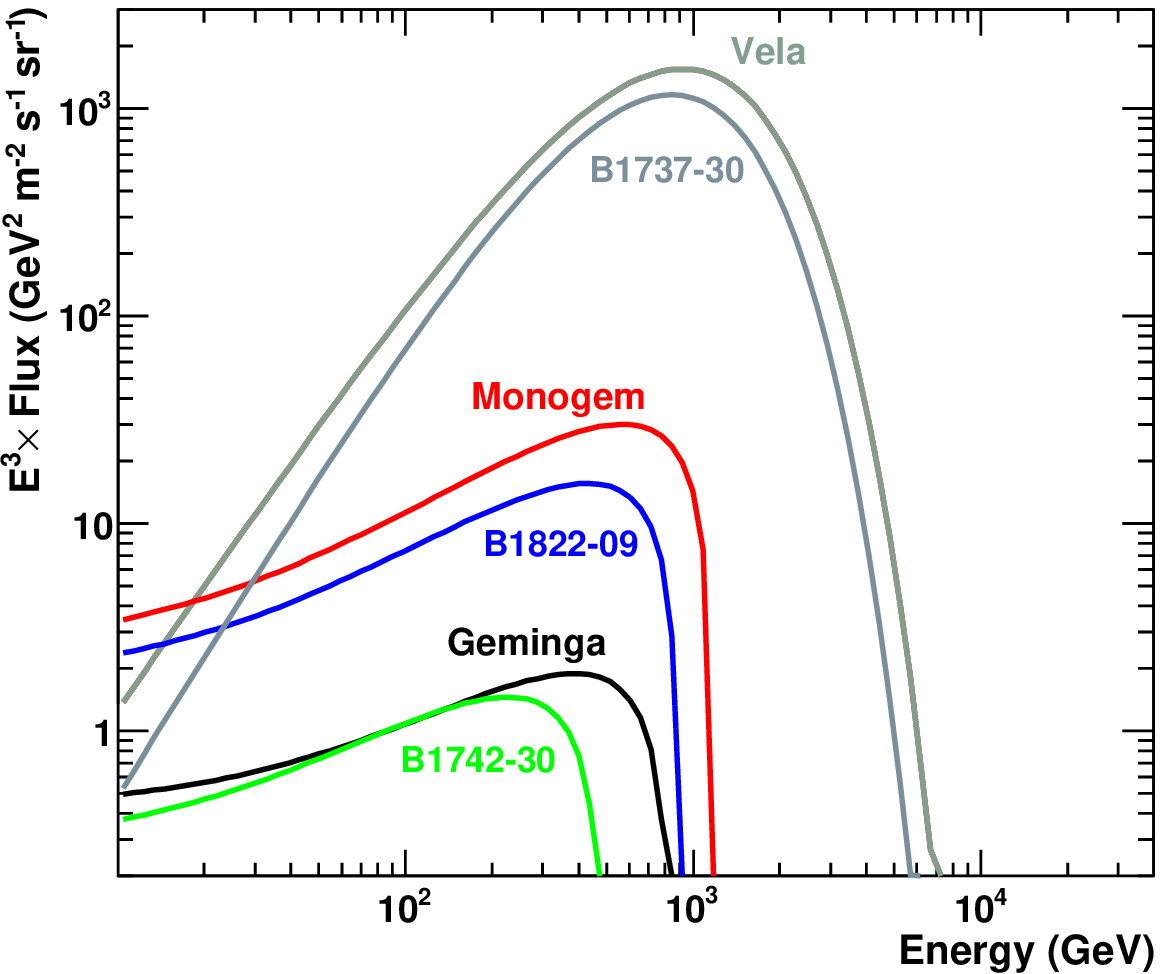}
    \caption{}
    \label{fig:comparison1}
  \end{subfigure}%
  ~
  \begin{subfigure}[t]{0.5\textwidth}
    \centering
    \includegraphics[width=0.80\textwidth]{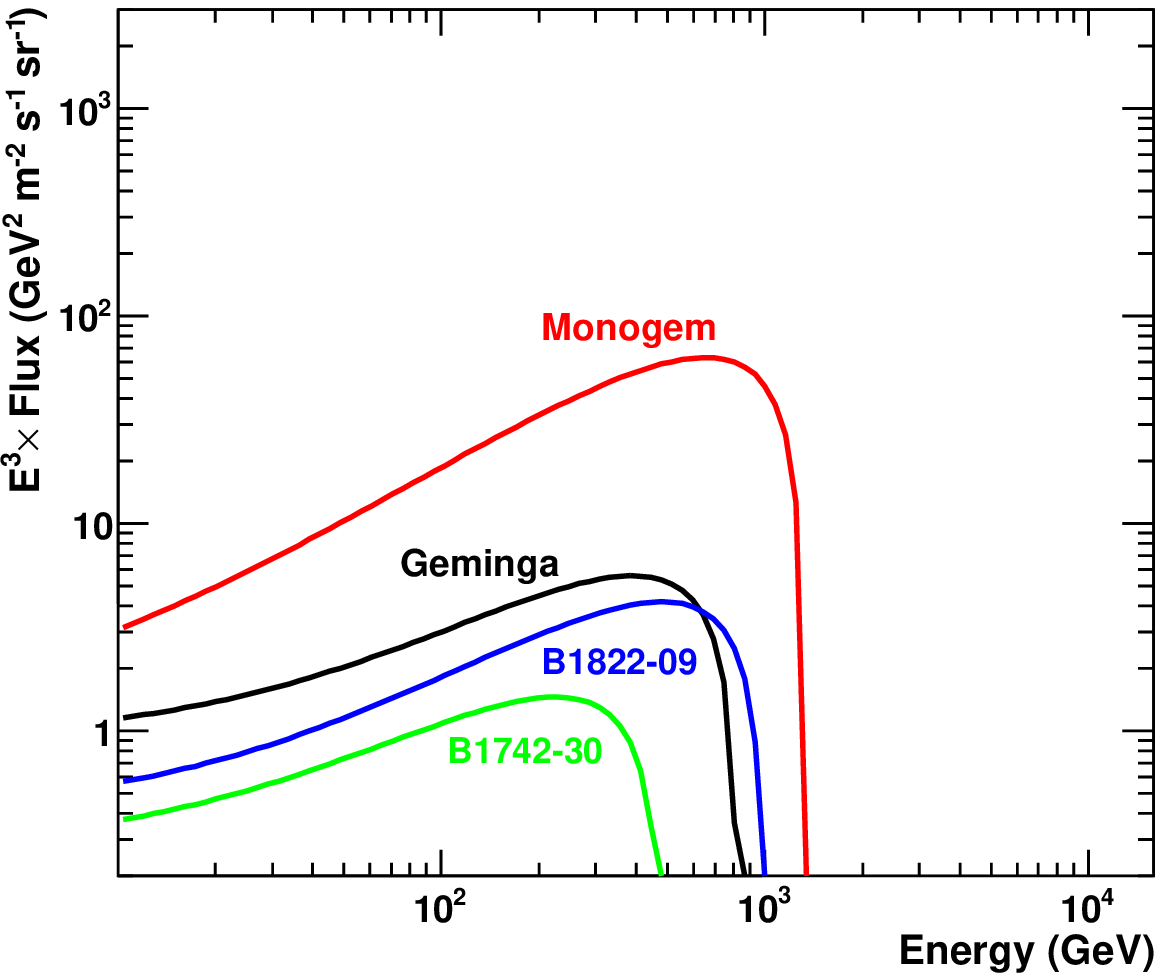}
    \caption{}
    \label{fig:comparison2}
  \end{subfigure}
  \caption{Comparison of the electron energy spectra from the candidate sources using a same set of the injection parameters ($\gamma = 2$, $E_\mathrm{cut} = 2$~TeV and $\eta W = 10^{46}$~eV) within the burst-like approximation (a) and when taking a time delay of 50~kyr before release (b).}
\label{fig:comparison}
\end{figure*}

As a next step of our analysis, we calculated the energy spectra of the observed cosmic-ray electrons from the most interesting objects using a reasonable combination of the free injection parameters ($\gamma = 2$, $E_\mathrm{cut} = 2$~TeV and $\eta W = 10^{46}$~eV). Although it is certainly unrealistic to presume that all sources share the same values of the injection parameters, this calculation allows us to have a better picture of the relative contributions from the different sources. As illustrated in figure~\ref{fig:comparison1}, when assuming the burst-like approximation, the contribution from Vela and B1737-30 outweigh all the others with the notable exception for low energies. In fact, a young source like Vela or B1737-30 mainly contributes at high energy because high-energy electrons are the first to reach Earth (see also figure~\ref{fig:time2}). The contributions from Geminga, B1822-09 and B1742-30 are all weaker than that of Monogem while the contributions from the other objects are utterly insignificant. Monogem contributes to the electron flux mainly at $\lesssim 20$~GeV. When we accept that electrons are confined in the neighborhood of the source a certain time ($\sim 50$~kyr) before release, the signals from Vela and B1737-30 just disappear (figure~\ref{fig:comparison2}). The situation for the other sources barely changes, with Monogem still predominant, but all the signals are somewhat enhanced except for B1822-09.

The steepening at high energy in the different energy spectra is determined by $E_\mathrm{cut}$ and the age of the source. Indeed, the maximum energy $E_\mathrm{max}$ achievable by a source of age $\tau_\mathrm{S}$ can be derived from relation~(\ref{eq:be}):
\begin{equation}
  E_\mathrm{max} \simeq \frac{1}{b_0 \tau_\mathrm{S}}.
\end{equation}

\noindent For Vela, $E_\mathrm{max} \simeq 24$~TeV, while for Monogem, Geminga and B1822-09, $E_\mathrm{max} \simeq 1$-2~TeV. The observation by HESS of a steepening in the cosmic-ray electron energy spectrum at $\sim 1$~TeV \cite{AHA08, AHA09} does not support a young source like Vela but rather favors middle-aged objects like Monogem, B1822-09 and Geminga.

\subsection{Electron flux and positron fraction}
\label{sec:flux}
Assuming a two-component model, we calculated the flux of cosmic-ray electrons (and the positron fraction) from Monogem. We then superimposed the contribution from this source on the otherwise featureless distant-source spectrum (background), represented here by the outcome of GALPROP. We used one of the GALPROP reference models that gives good fits to cosmic-ray data, namely the plain diffusion model \cite{STR98, PTU06}.

GALPROP is a well-known numerical package for calculating the propagation of cosmic rays (not only electrons) in a realistic distribution of matter. It gives the local interstellar spectra of the different cosmic-ray species by solving numerically the diffusion-loss equations, such as (\ref{eq:transport}). It is the most representative example of conventional models. As already mentioned, in these models it is presumed that equal amounts of secondary positrons and negative electrons are created in collisions between cosmic-ray nuclei and the interstellar medium, in addition to negative electrons directly produced at the same astrophysical sources as cosmic-ray nuclei. Although conventional models fail to match HECRE data, GALPROP is eminently suitable for the background estimation.

\begin{figure*}[t!]
  \centering
  \begin{subfigure}[t]{0.5\textwidth}
    \centering
    \includegraphics[width=0.9\textwidth]{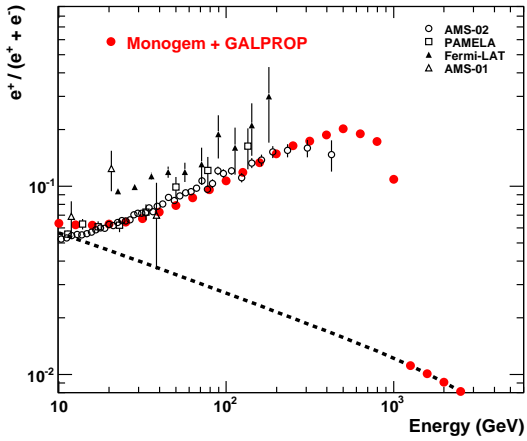}
    \caption{Positron fraction.}
    \label{fig:positron_fraction}
  \end{subfigure}%
  ~
  \begin{subfigure}[t]{0.5\textwidth}
    \centering
    \includegraphics[width=0.9\textwidth]{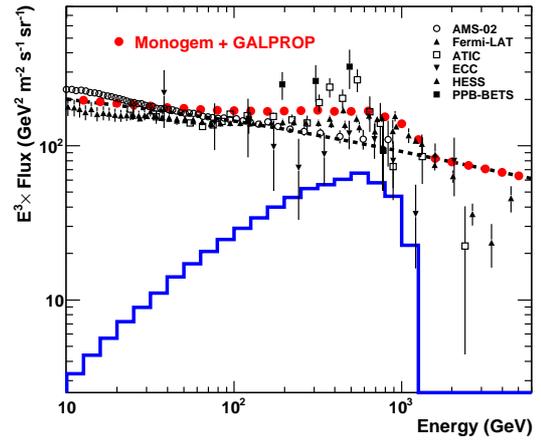}
    \caption{Total e$^+$+e$^-$ flux.}
    \label{fig:all_electrons}
  \end{subfigure}%
  \vspace{4ex}
  \begin{subfigure}[t]{0.5\textwidth}
    \centering
    \includegraphics[width=0.9\textwidth]{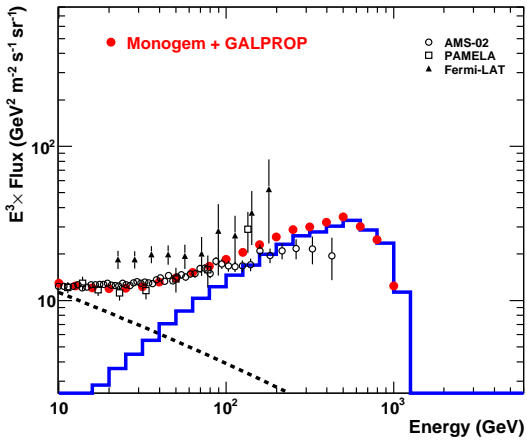}
    \caption{Only e$^+$ flux.}
    \label{fig:positrons}
  \end{subfigure}%
  ~
  \begin{subfigure}[t]{0.5\textwidth}
    \centering
    \includegraphics[width=0.9\textwidth]{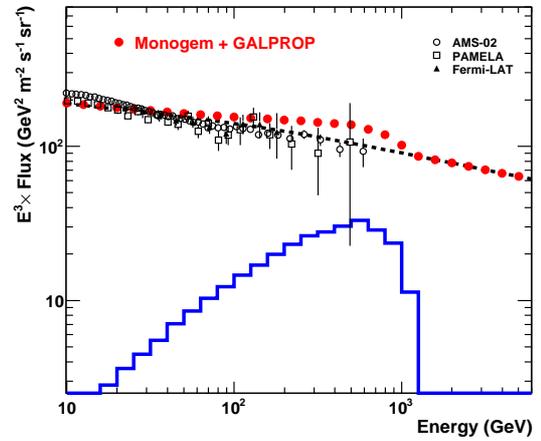}
    \caption{Only e$^-$ flux.}
    \label{fig:negative_electrons}
  \end{subfigure}
  \caption{Positron fraction and flux of cosmic-ray electrons as a function of energy. The dashed line represents the background (GALPROP) and the histogram in (b), (c) and (d) the contribution from Monogem. ``Monogem + GALPROP'' is intended to mean Monogem contribution on top of the background. Also shown are the experimental data from recent measurements: AMS-02 \cite{ACC14, AGU14b, AGU14a}, PAMELA \cite{ADR13, ADR11}, Fermi-LAT \cite{ACK12, ACK10b}, AMS-01 \cite{AGU07}, ATIC \cite{CHA08}, ECC \cite{KOB12}, HESS \cite{AHA08, AHA09} and PPB-BETS \cite{YOS08}.}
  \label{fig:flux}
\end{figure*}

As can seen in figure~\ref{fig:flux}, the contribution from Monogem reproduces simultaneously and satisfactorily all the experimental data: the increase in the positron fraction (\ref{fig:positron_fraction}), the hardening observed in the cosmic-ray electron spectrum (\ref{fig:all_electrons}), as well as the negative electron (\ref{fig:positrons}) and the positron spectra (\ref{fig:negative_electrons}) separately. It should be stressed here that we used in all these calculations the same combination of the free injection parameters ($\gamma = 2$, $E_\mathrm{cut} = 2$~TeV, and $\eta W = 4 \times 10^{46}$~erg). The other sources, like B1822-09 and Geminga, also reproduce the observed spectra after, of course, tuning the injection parameters. Note as well that the traditional methods lead to similar results as shown by many previous studies (see, e.g., \cite{HOO09, MAL09, YUK09, DEL10, BLA11, PRO11, DIM14}). The novelty here, as aforementioned, is that we used a pure Monte Carlo treatment of the propagation of HECREs from nearby single sources, instead of solving transport equations.

\subsection{Anisotropy}
\label{sec:anisotropy}
The anisotropy in the arrival directions of cosmic-ray electrons is thought to be the smoking gun for the astrophysical scenario. High-energy electrons from local sources would indeed experience little deflection by the galactic magnetic field and the distribution of their arrival directions should then show significant anisotropy. Likewise, the observation from any direction in the sky of an excess of cosmic-ray electrons with energies high enough to minimize both geomagnetic and heliospheric effects could potentially be the signature of such nearby sources.

The benchmark method for studying anisotropy involves a multipole expansion of the electron intensity fluctuations:
\begin{equation}
  \frac{I_{\mathrm e}(\theta,\, \phi) - \langle I_{\mathrm e} \rangle}{\langle I_{\mathrm e} \rangle} = 
	  \sum_{l=0}^{\infty} \sum_{m=-l}^{l} a_{lm} Y_{lm}(\pi/2-\theta, \, \phi),
\end{equation}
\noindent where $I_{\mathrm e}(\theta,\, \phi)$ is the cosmic-ray electron intensity in the direction defined by the polar angles ($\theta,\, \phi$) and $\langle I_{\mathrm e} \rangle$ the average value over the whole sky map. The functions $Y_{lm}$ are the spherical harmonics and $a_{lm}$ their corresponding weights. The coefficients of the angular power spectrum of the fluctuations are defined as
\begin{equation}
  C_l = \frac{1}{2l+1} \sum_{m=-l}^{l} |a_{lm}|^2.
\end{equation}
The amplitude of dipole anisotropy is then
\begin{equation}
  \delta = 3 \sqrt{\frac{C_1}{4\pi}}.
\end{equation}

Many theoretical studies show that, even after the diffusive propagation of cosmic-ray electrons, a small dipole anisotropy should be present in the direction of the dominant nearby source at sufficiently high energies \cite{SHE71, PTU95, BUS08}. However, the data from AMS-02, Fermi-LAT and PAMELA are found to be compatible with isotropy at any angular scale \cite{AGU13, ACC14, ACK10a, ADR15}.

The assessment of the degree of anisotropy induced by candidate sources can be handled very efficiently with our Monte Carlo model. That is because we have the arrival direction of each observed electron, besides its energy and lifetime. We analyzed the distribution of arrival directions by using the spherical harmonic method in galactic coordinates within the framework of HEALPix\footnote{http://healpix.sourceforge.net/} \cite{GOR05}. This software is the standard algorithm for pixelization on the sphere, producing a subdivision of a spherical surface in which each pixel covers the same surface area (or solid angle). For all our sky maps we used a grid of 12,288 pixels corresponding to an angular resolution of about $2^\circ$ (angle subtending the centers of 2 adjacent pixels). To maximize the appearance of anisotropy we replaced the content of each pixel with the integrated number of events of all its closest neighbors, the integration extending up to a maximum angular radius defined by the considered angular scale \cite{ACK10a, ADR15}. Such correlated maps enhance sensitivity to week anisotropic signals, especially those distributed through multiple adjacent bins.

\begin{figure*}[t!]
  \centering
  \begin{subfigure}[b]{0.5\textwidth}
    \centering
    \includegraphics[width=0.9\textwidth]{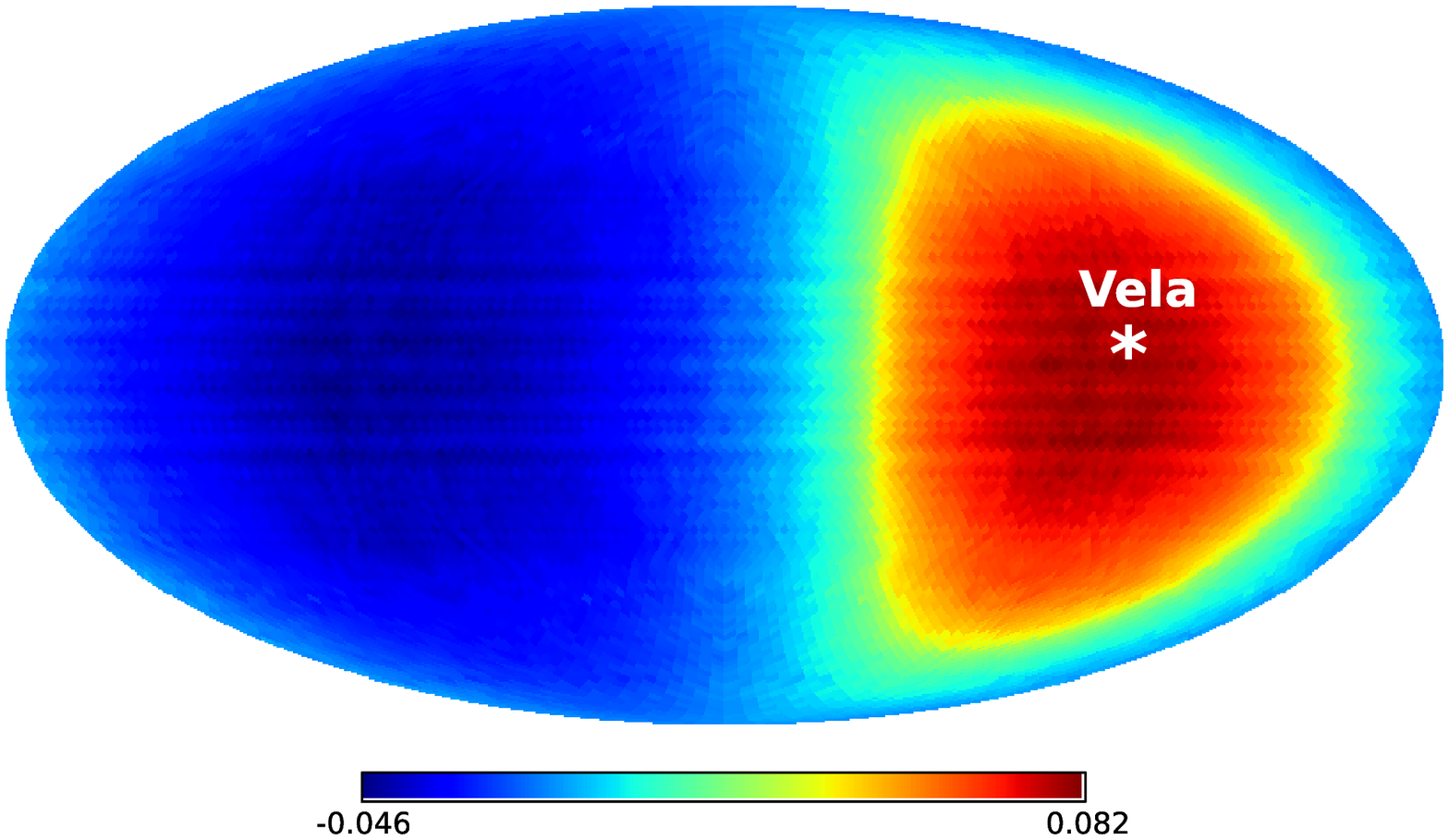}
  \end{subfigure}%
  ~
  \begin{subfigure}[b]{0.5\textwidth}
    \centering
    \includegraphics[width=0.9\textwidth]{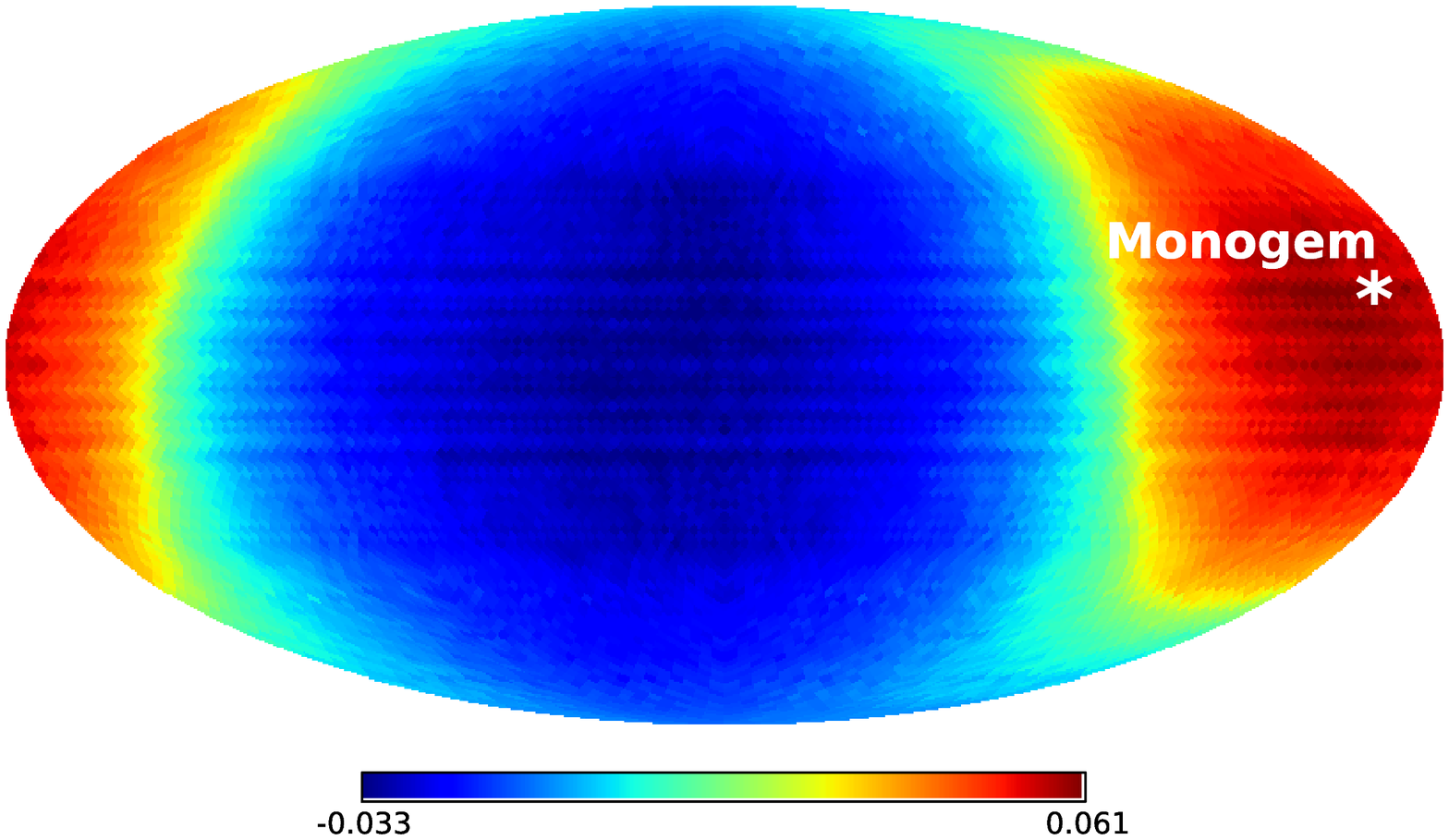}
  \end{subfigure}
  \caption{Mollweide views of the dipole anisotropy induced by Vela (left) and Monogem (right) for cosmic-ray electrons with energies above 60~GeV at a $60^\circ$ angular scale.}
  \label{fig:anisotropy}
\end{figure*}

We started our analysis by assessing anisotropy in the case of a perfectly isotropic sky. For a set of $10^4$ samples of isotropic cosmic-ray electrons, similarly sized ($10^5$), we obtained an average dipole anisotropy $\langle \delta \rangle = 0.9\%$ with a standard deviation $\sigma = 0.4\%$. Afterwards we investigated the anisotropy induced by the potential sources for electrons with energies above 60~GeV (Table~\ref{tab:anisotropy}). Mollweide views for Vela and Monogem are illustrated in figure~\ref{fig:anisotropy}. We made the calculations in case of the complete absence of any background and in case of the existence of an overwhelming isotropic background (10 times the source signal). As a matter of fact, looking closely at figure~\ref{fig:all_electrons}, it can be seen that the background at 60~GeV is approximately one order of magnitude higher that the contribution from Monogem. This isotropic background may reflect the overall effect of more distant sources, as hinted by the very long tail of the lifetime distributions (see figure~\ref{fig:time}). We used for all sources the same combination of the free injection parameters ($\gamma = 2$ and $E_\mathrm{cut} = 2$~TeV). We also considered a time delay of 50~kyr with respect to the source birth except for Vela. We set the angular scale at $60^\circ$ and the sample size at $10^5$ in each case. Besides the amplitude of dipole anisotropy $\delta$, we calculated the statistical significance of the electron excess, defined here as
\begin{equation}
  S = \frac{I_{\mathrm e}^{\mathrm max} - \langle I_{\mathrm e} \rangle}{\sigma}.
\end{equation}
\noindent $I_{\mathrm e}^{\mathrm max}$ is the electron intensity from the hottest pixel. This pixel does not necessarily coincide with the source pixel but always lies very close.

\begin{table}[tbp]
  \centering
  \begin{tabular}{|l|c|c|c|c|c|}
    \hline
     &  Vela & Monogem & B1822-30 & Geminga\\ 
    \hline
    $\delta$~(\%) & 6.2 (67.8) & 4.6 (47.9) & 3.5 (40.3) & 3.1 (36.1) \\
    \hline
    $S$~($\sigma$) & 2.15 (2.13) & 2.21 (2.15) & 2.25 (2.13) & 2.32 (2.14) \\
    \hline
  \end{tabular}
  \caption{Amplitude of the dipole anisotropy ($\delta$) and statistical significance of the electron excess ($S$)  for the candidate sources in case of the existence of a background one order of magnitude higher than the source signal and, in brackets, in case of the absence of any background.}
  \label{tab:anisotropy}
\end{table}

These calculations show in the first place that in no case the electron excess is statistically significant, $S$ always being very low ($\approx 2 \sigma$). This result agrees with experimental data \cite{AGU13, ACC14, ACK10a, ADR15}. However, the obtained amplitudes of dipole anisotropy are in no way consistent with Fermi data, which set the upper limit of $\delta$ above 60~GeV at $0.5\%$ \cite{ACK10a}. The discrepancy is stronger for the young source Vela than for the middle-aged sources Monogem, Geminga and B1822-30. Admittedly, the overwhelming background strongly reduces the dipole anisotropy, but not enough to be in close agreement with experimental data. The idea that the energy spectrum of cosmic-ray electrons is dominated at high energy by one single source is definitely not supported by the observations. Excluding Vela-like sources, there are at least three objects that may play a crucial role, namely Monogem, B1822-30 and Geminga (see figure~\ref{fig:comparison2}). While Monogem and Geminga are located very close together on the sky dome, B1822-30 is roughly diametrically opposite, which would cancel out the anisotropy caused by the two first objects. Indeed, if we neglect the contribution from Geminga and assume that only Monogem and B1822-09 prevail at high energy, each contributing up to 50\%, the dipole anisotropy drops to 6\% in the absence of the background and 0.6\% in its presence. This last value is more in line with Fermi data. In sum, the non-observation of anisotropy, which challenges the astrophysical scenario, can be understood only if we presuppose the existence at high energy of at least two dominant sources in such a configuration that one nullifies the effect of the other, in addition to a significant background induced by more distant sources.

\section{Conclusion}
\label{sec:conclusion}

Regarding the spectral peculiarities observed at high energy for cosmic-ray electrons, which still resist a unified interpretation, there is now a manifest need to explore new avenues in the hope of solving this puzzle. Most often the way of tackling this problem revolves around the resolution of the transport equation describing the galactic propagation of these particles. This work demonstrates the feasibility and the relevance of fully three-dimensional time-dependent Monte Carlo simulation, which can supplement the reference method by providing additional information on the electron-by-electron fluctuations. This approach is proving to be particularly efficient at investigating more deeply the possible astrophysical origin of HECREs from nearby single sources and its benefit appears plainly in addressing the anisotropy issue. The proposed algorithm is quite simple, very flexible and highly scalable.

When looking at the lifetime and energy distributions of observed cosmic-ray electrons, we came to a first list of candidate sources that includes Vela, B1737-30, Monogem, B1822-09, Geminga and B1742-30. When assuming the burst-like approximation, we found out that the signal from a young source like Vela would outweigh  all other signals and would produce an exceedingly large anisotropy amplitude, which obviously disagrees with observations. Considering a time delay of 50~kyr with respect to the source birth, we obtained a new list of candidate sources with the middle-aged Monogem taking the lead. We showed then that Monogem is able to reproduce simultaneously all the experimental data, namely the e$^+$/(e$^-$+e$^+$), e$^-$+e$^+$, e$^-$ and e$^+$ energy spectra. However, Monogem is also in conflict with the upper bounds of dipole anisotropy set by Fermi. These calculations show, in fact, that the non-observation of anisotropy does not support the thesis of one single object dominating at high energy, even with the existence of an overwhelming background. But these calculations also indicate that there are other key players, specifically B1822-09 and Geminga. While Monogem and Geminga are located close together on the sky dome, B1822-09 is situated just in the opposite side. In this way, B1822-09 could very well negate the possible anisotropy caused by Monogem and Geminga. But to be in accordance with Fermi data, we had to consider in addition a significant isotropic background likely to arise from distant sources.

Last but not least, we are aware of the shortcomings of this new software. The main line of thinking of this work is to build a workable program with a minimum version to save computing time, and make improvements afterwards. Future refinements will focus on the propagation of HECREs through more realistic magnetic field structures.

\acknowledgments
We gratefully acknowledge computing support generously provided by the Center for High Performance Computing (CHPC) in Cape Town (South Africa), and by the Research Center on Scientific and Technical Information (CERIST) in Algiers (Algeria) through the HPC platform Ibnbadis. We would also like to thank the reviewers for their relevant comments, resulting in a notable improvement of the paper.

\bibliography{hecre}

\end{document}